\documentclass[graybox]{svmult}
\pdfoutput=1

\usepackage{type1cm}        
\usepackage{graphicx}        
\usepackage{multicol}        
\usepackage[bottom]{footmisc}

\usepackage{newtxtext}       %
\usepackage{newtxmath}       

\usepackage[utf8]{inputenc}
\usepackage{amsfonts,amsmath}

\def\R{\mathcal{R}}

\date{September 2020}

\begin{document}

\title*{Describing, modelling and forecasting the spatial and temporal spread of COVID-19 -- A short review}
\titlerunning{Spatio-temporal spread of COVID-19}
\author{Julien Arino}
\institute{Julien Arino \at Department of Mathematics, University of Manitoba, Winnipeg, Manitoba, Canada, \email{Julien.Arino@umanitoba.ca}}
%
%
\maketitle

\abstract{SARS-CoV-2 started propagating worldwide in January 2020 and has now reached virtually all communities on the planet. 
This short review provides evidence of this spread and documents modelling efforts undertaken to understand and forecast it, including a short section about the new variants that emerged in late 2020.}

\section{Introduction}
On 21 February 2003, a disease now known as the severe acute respiratory syndrome (SARS) arrived in Hong Kong when a physician from Guandong Province in Mainland China bearing the SARS coronavirus (SARS-CoV), checked in at the Metropole Hotel \cite{Cherry2004}. The primary case in Hong Kong was by no means the index case: the virus had been circulating in Guandong since at least November 2002 \cite{ChanYeungXu2003}. However, that patient triggered a chain of infections that, together with earlier cases in China, led to 8,098 known cases and 774 deaths in 28 countries \cite{BowenLaroe2006} and was declared a pandemic by the World Health Organisation (WHO).

Similarly, it is not certain at the time of writing that SARS-CoV-2 and its associated disease COVID-19 had its index case in Wuhan, Hubei Provice, China. What is certain, on the other hand, is that it is in Wuhan that COVID-19 underwent its first noticeable amplification phase, following which it spread rapidly across the world, to the point that there are now very few top level jurisdictions not having reported COVID-19 cases.

SARS-CoV-2 is the third novel Coronavirus to emerge in the 21st century (after SARS and the Middle East respiratory syndrome -- MERS \cite{AzharLaniniIppolitoZumla2017}) and the second to generate a pandemic (a third pandemic was triggered by the H1N1 influenza outbreak in 2009). COVID-19 is also the most devastating pandemic in over a century in terms of its death toll as well as its economic and societal impact.

Here, I review some aspects of the spatio-temporal spread of COVID-19. 
Some caveats are in order. 
Firstly, while spatial epidemiology is not the most popular topic among modellers, it does remain a vast field where a myriad of approaches coexist; surveying the work done on the topic would require an entire monograph. 
While I have tried to be inclusive, it is certain that I have omitted some topics or techniques. 
I am for instance making the choice to describe mostly mechanistic models of spread, be they deterministic or stochastic, mathematical or computational. 
Some very good statistical work has been published on the topic of COVID-19, but I focus here to a large extent on models that can explain reality perhaps at the detriment of forecasting power. 
Secondly, I am aware that many modellers who have worked or are working with public health authorities may not have had the time yet to publish their work.
Except for my work, I report here only on papers already published or available on recognised preprint servers.
Thirdly, new variants of concern (VoC) were detected while this paper was under review. 
A short section at the end of this document to describe what little is known in terms of spatio-temporal spread of these variants.

Finally, even though some work makes use of data at a very fine spatial resolution, in keeping with the philosophy of some prior work \cite{Arino2020a}, I focus on models that can be used with publicly available data.

This review is organised as follows. 
First, in Section~\ref{sec:mechanisms}, I provide an overview of the mechanisms that lead to the spatial spread of infections and three of the major types of models that have been used to study it. 
In Section~\ref{sec:chronology_and_characteristics}, I then describe the spread of COVID-19 from a chronological point of view. 
Finally, in  Section~\ref{sec:covid19_modelling}, I discuss modelling work specific to COVID-19.

\section{Spatialised infections -- mechanisms and models}
\label{sec:mechanisms}
Before considering work specific to COVID-19, let me spend some time on the spatialisation of infectious diseases in general. 
Indeed, while COVID-19 presents specific challenges, it is by no means the first spatial epidemic that humanity is confronted to; for instance, a simple description of the spatio-temporal trajectory of the Plague of Athens can be found in the History of the Peloponnesian War \cite{Thucydides430BCE}, which was written almost 2,500 years ago; on a more local scale, spatial epidemiology can be traced to the cholera epidemic of London in 1854 \cite{Snow1855}.
There is therefore much understanding to be gained about the current crisis by considering what was known prior to its start.

\subsection{How does an infectious disease become spatial?}
\label{subsec:spatialisation}
Different conceptual models explain the mechanisms that lead to the spatialisation of an infectious disease, leading to potentially different modelling paradigms.

Working at the level of the individual, one can envision spatial spread as the repetition of inter-individual spread events. Individuals are mobile in space and it is their movement while bearing the infectious pathogen that leads to the disease becoming spatial, when they come into contact with susceptible individuals who are also mobile.
This description falls mostly into what have been called Markovian contact processes \cite{Mollison1978}. 
When considered at the population level, this leads to models using partial differential equations and is particularly appropriate for describing the spread of a disease where the hosts can move freely, such as epizooties. Such a description can also lead to network or agent-based models.

Work with my collaborators usually instead focuses on locations and adopts a vision of spatial spread articulated in \cite{ArinoBajeuxPortetWatmough2020}: an infectious disease becomes spatialised by the repetition of processes summarised as \emph{importation}, \emph{amplification}, \emph{exportation} and \emph{transport}. Importation itself is the event when an individual infected with the disease reaches a new location. 
Importation is \emph{successful} if the imported case leads to at least one local transmission event. 
This way of thinking about spread is easy to reconcile with data, since locations are jurisdictions in the context of public health.
It also matches the \emph{cones of resolution} that some geographers use when they think about the spatial spread of epidemics; see \cite{AnguloTakigutiPederneirasCarvalhodeSouzaEtAl1979} and references therein.
See also \cite{HuGongSunZhou2013,HuGongZhouSunEtAl2013}, which consider the roles of the different levels of mobility on the spread of SARS in and to and from Beijing.

\subsection{What are the main drivers of spatial spread?}
\label{subsec:drivers_of_spatial_spread}
Whatever the way one conceptualises the spatialisation process, the main driver of spatial spread is human mobility. Long range fast movements using air travel have considerably changed the way diseases spread and while amplification in a location remains driven by population effects, the initial spread is to a large extent driven by air travel. This was shown for SARS \cite{BowenLaroe2006}, the 2009 H1N1 influenza pandemic \cite{KhanArinoHuRaposoEtAl2009} and MERS \cite{GardnerChughtaiMacIntyre2016}, for instance. 
Long distance high speed train travel has also been associated to spread; see, e.g., \cite{CaoZengZhengWangEtAl2010}.
It is interesting to note, though, that despite the highly heterogeneous nature of spatio-temporal spread brought on by modern travel modalities, continental-level effects can still be observed \cite{GeogheganSaavedraDucheneSullivanEtAl2018}.

\subsection{How does one model a spatialised infectious disease?}
\label{subsec:spatial_models}

As mentioned in the Introduction, I focus here on mechanistic models. 
There are many ways to model the spatio-temporal spread of infectious diseases. 
Let me present the main contenders; see an interesting and more complete list in \cite{Robertson2019} or \cite{MollisonKuulasmaa1985}.
I do not detail reaction-diffusion equations, because, to the best of my knowledge, they have seen very little use in modelling the spatio-temporal spread of COVID-19; readers are referred to \cite{RassRadcliffe2003}, for instance, for more details on deterministic aspects involving such systems.
See \cite{Mollison1977} for a seminal review of the link between stochastic and deterministic spatial models, as well as interesting overviews in \cite{ColizzaBarthelemyBarratVespignani2007}.

\subsubsection{Agent-based models} 
Agent-based models (ABM) consider populations of autonomous \emph{agents} that interact following some rules \cite{Epstein2006}. Agents have a set of characteristics that can be modified through their interactions with other agents. Although there are some attempts to mathematise some of the properties of such systems, they remain for the most part computational tools that need to be studied using a large number of simulations. Their strength lies in their realism: an agent can be given realistic behavioural characteristics (schedule, place of residence or of work, etc.).
ABM are also easier to implement because they require very little mathematical background.
Agent-based models have proved most useful when considering the effect of individual behaviours on the spread of infection in smaller populations. 
For instance, they have been used to study individual protective behaviour \cite{KarimiSchmittAkgunduz2015}, the effect of presenteism while infected with a disease \cite{KumarGrefenstetteGallowayAlbertEtAl2013}, the risk in small isolated communities \cite{CarpenterSattenspiel2009,LaskowskiDuvvuriBuckeridgeWuEtAl2013}, the effect of social distancing \cite{SinghSarkhelKangMaratheEtAl2019} or the role of avoidance behaviour when vaccines have low effectiveness \cite{VilchesJaberiDourakiMoghadas2019}.
Examples of spatialised problems (all about influenza) that were studied using agent-based models include its spread in slums of Delhi \cite{AdigaChuEubankKuhlmanEtAl2018}, the use of a hybrid approach involving networks to describe the social structure and ABM to describe inter-individual spread in Forsyth Country, NC \cite{GuoLiPetersSnivelyEtAl2015}, the potential for social structure to generate inequalities in incidence in different areas of a county \cite{KumarPiperGallowayHadlerEtAl2015} or the spread within an airport terminal \cite{ShaoJia2015}. 
See also \cite{OhkusaSugawara2007,OhkusaSugawara2009}, which use a detailed location survey to conduct a simulation of the spread of influenza in Japan.
ABM are also useful as a means to model evolution; see, for instance, \cite{GrietteRaoulGandon2015}, where an ABM is used to model evolution of virulence at the front line of a spreading epidemic

The area where ABM have proved most informative is when considering spread of infections within areas where movement is constrained and generalised contact is impossible, such as buildings or cruise ships. For instance, when considering nosocomial infections, it is possible to monitor health care personnel movement and use this data to parametrise an ABM of spread within a hospital \cite{HornbeckNaylorSegreThomasEtAl2012}, or to formulate a model of spread between beds within an intensive care unit \cite{HornbeckNaylorSegreThomasEtAl2012}.

However, ABM lose in value when populations become larger, except in rare instances where unexpected \emph{emergent} behaviour occurs.
Where the law of large number applies, it is indeed less computationally onerous to use ``classic'' deterministic or stochastic models. For instance, the model in \cite{ArinoBrauerDriesscheWatmoughEtAl2006} reproduces almost exactly the behaviour of the agent-based model in \cite{LonginiNizamXuUngchusakEtAl2005}, but furthermore gives access to explicit expressions of the basic reproduction number $\R_0$ and the final size of the epidemic. See also the comparisons in \cite{AjelliGoncalvesBalcanColizzaEtAl2010,DalgicOezaltinCiccotelliErenay2017}.

Altogether, agent-based models are powerful tools of investigation at the hyperlocal scale. Considering agents consisting of groups of individuals instead of individuals also allows to operate at a higher spatial scale, although models then lose some of the interesting properties they have at the finer scale.


\subsubsection{Network models}
\label{subsec:network_models}
Network models are very similar to agent-based models, of which they were, essentially, the inspiration. 
The two types of models are sometimes difficult to distinguish.
In epidemiology, work on networks and ABM was popularised in particular by the NIGMS Models of Infectious Disease Agent Study, which led for instance to EpiSimS \cite{ChowellHymanEubankCastilloChavez2003,Eubank2005}.
In network models, \emph{nodes} are typically simpler than agents in agent-based models; the most straightforward example would be a network consisting of nodes (individuals) that can be in two states: susceptible to the disease or infected and infectious with it.
If there is an edge in the network between two nodes, this means the two nodes came into contact; in the case of a network used to model disease spread, this indicates that a contact took place, which could lead to the transmission of the disease.
One promising direction of research that has been explored using networks is that of the link between network structure and shape of the epidemic curve; see, e.g., \cite{Carnegie2018,ChowellSattenspielBansalViboud2016}.
This is particularly important during the early spread of a disease and has been considered in a variety of contexts using network models. 

Because they are simpler than ABM, network models are more amenable to analysis; see, e.g., \cite{BarnardBerthouzeSimonKiss2019,BifolchiDeardonFeng2013}. 
Originally, tools used to study the dynamics of network epidemic models originated in statistical mechanics
\cite{MeyersNewmanPourbohloul2006,MeyersPourbohloulNewmanSkowronskiEtAl2005}.
Because networks allow to incorporate a more realistic description of the contact process while maintaining some level of analytic tractability, comparing their dynamics with that of classical models is useful. In \cite{AlexanderKobes2011}, this is done for instance for a two-strains influenza model with vaccination. 
\cite{PourbohloulAhuedDavoudiMezaEtAl2009}
There has been a move lately towards characterising the dynamics of smaller networks using the properties of individual nodes rather than through distributions of these properties; see, e.g., \cite{BallHouse2017}.

While examples of use of network models in mathematical epidemiology abound, their use in situations that are specifically spatial are not as common. Instances include \cite{FangChenHu2005}, who considered an SEIR model set in a lattice and simulated using a Monte Carlo process, to incorporate both stochasticity and space, \cite{BarmakDorsoOteroSolari2011}, who consider the spread of dengue in city blocks or \cite{FirestoneWardChristleyDhand2011}, who consider the spread of equine influenza. The latter paper illustrates the strength of the method, in that they have access to extensive data on horse movement between locations and are able to assess the effect of the topology of the networks, both for long range movement and shorter contact patterns with locations, on the spread of the infection.

Networks are also a natural candidate for considering the spread of infections using the air transportation network as a conduit. This was done for instance with SARS \cite{BowenLaroe2006}.

\subsubsection{Metapopulation models}
Also known as patch models, metapopulations couple together (typically similar) models, with each model encoding for the dynamics of the disease in a population and coupling representing the movement of individuals between the populations \cite{Arino2017}, the average time spent in remote locations  \cite{BicharaIggidr2018} or the interactions between populations.
Metapopulation models had been used computationally since the early 1970s, for instance to consider the spread of influenza within countries \cite{FlahaultLetraitBlinHazoutEtAl1988,RvachevLonginiJr1985}. 
They have known a resurgence since the beginning of the 21st century, because, on the one hand, computing resources made simulating them easier and, on the other hand, papers such as  \cite{ArinoVdD2003a,ArinoDavisHartleyJordanEtAl2005} showed that linear algebra techniques could be used to render the study of such systems very similar to that of their constituting systems.
See \cite{BallBrittonHouseIshamEtAl2015} for a list of problems the authors identify as interesting challenges in the field.

Metapopulations are now quite popular and have been used in a variety of settings. A lot of work concerns investigation of properties of spatial models. For instance, with coauthors, I investigated the effect of lowering travel rates between locations \cite{ArinoJordanDriessche2007} and of interconnection between a large urban centre and smaller satellite cities \cite{ArinoPortet2015}.
Other interesting issues studied include the effect of vaccination targeted at high risk areas \cite{AzmanLessler2015}, cooperation between governments on vaccination policy \cite{KlepacMegiddoGrenfellLaxminarayan2016}.
Geographically targeted vaccination has also been considered at smaller spatial scales; see, e.g., \cite{AsanoGrossLenhartReal2008,AzmanLessler2015,HallEganBarrassGaniEtAl2007,KellyTienEisenbergLenhart2016,KhatuaKarNandiJanaEtAl2020,MatrajtHalloranLongini2013}. 
Other spatial control issues have been considered in \cite{BockJayathunga2019,GaythorpeAdams2016,GlassBarnes2013,HarvimZhangGeorgescuZhang2019,KimLeeLeeLee2017,LeeCastilloChavez2015,MatthewsHaydonShawChaseToppingEtAl2003}.

Papers addressing issues that are present also with COVID-19 have considered infection during transport \cite{ArinoSunYang2016}, in particular in relation to entry screening \cite{LiuTakeuchi2006} as well as exit and entry screening \cite{LiuChenTakeuchi2011}. Exit and entry screening were also considered in \cite{WangLiuWangZhang2015}. Some work has also considered the effect of media-induced social distancing \cite{GaoRuan2011,SunYangArinoKhan2011}.

Metapopulation models were used to consider specific diseases as well; the spread of SARS \cite{RuanWangLevin2006}, age-structured contact patterns during the 2009 H1N1 pandemic \cite{ApolloniPolettoColizza2013}, chikungunya \cite{ChadsuthiAlthouseIamsirithawornTriampoEtAl2018}, dengue \cite{MishraAmbrosioGakkharAzizAlaoui2018}, cholera \cite{EisenbergShuaiTienDriessche2013} or malaria \cite{ArinoDucrotZongo2012,GaoRuan2012,GaoDriesscheCosner2019}; see also \cite{Riley2007}.

Standard metapopulation models are not well suited to consider the hyperlocal scale, because they assume homogeneity within the constituting units. 
In \cite{ApolloniPolettoColizza2013}, an interesting approach is used that allows more heterogeneous contacts within patches. 
Similarly, in \cite{MossNaghizadeTomkoGeard2019}, the authors consider behaviour at the hyperlocal scale but still within a metapopulation model.
In \cite{AjelliGoncalvesBalcanColizzaEtAl2010}, the behaviour of an ABM is compared with that of a stochastic metapopulation model.

\section{Chronology and characteristics of COVID-19 spread}
\label{sec:chronology_and_characteristics}
To describe the spatio-temporal spread of COVID-19, I use the previously discussed framework of \cite{ArinoBajeuxPortetWatmough2020}. 
There is some discrepancy in reporting units, but to some extent, one can think along the lines of the ISO 3166 standard \cite{ISO2013}. \emph{Global} spread occurs between ISO 3166-1 codes (countries, dependent territories and special areas of geographical interest). \emph{Local} spread within ISO 3166-1 codes occurs between ISO 3166-2 codes (provinces in Australia and Canada, d\'epartements in France, states in Brazil and the USA, etc.). Many countries also report at a finer geographical scale, which I still call \emph{local}: counties, regional health authorities or cities.
Anything below the city level is \emph{hyperlocal}; typically, this corresponds to commuting to work, school or shopping, but can even be mobility within a building.
In keeping with my avowed preference for publicly available data, my description involves higher level jurisdictions rather than spread at the hyperlocal scale, which is typically associated to confidential data. 

Throughout the description that follows, one should bear in mind that data used to describe the spread is very likely wrong in some instances, or rather, some jurisdictions might be reporting with a delay because of a lower capacity to detect cases. See for instance \cite{HaiderYavlinskyChangHasanEtAl2020}, in which two health security indices, the Global Health Security Index and the Joint External Evaluation, are used to assess the likelihood that countries detected COVID-19 early. (The work also shows that countries with higher values of these indices also saw reduced mortality from the disease to 1 July 2020, although this is likely not true anymore.)

\subsection{Chronology and characteristics of global spread}
\label{subsec:global_spread}
There is evidence that COVID-19 could have started its global spread in December 2019, with reports of a case in France \cite{DeslandesBertiTandjaouiLambotteAllouiEtAl2020} as well as suspicious cases \cite{BirtoloMaestriniSeverinoChimentiEtAl2020} and positive wastewater samples \cite{LaRosaManciniBonannoFerraroVeneriEtAl2020} in Italy. 
However, these retrospective analyses have yet to be confirmed, so at the time of writing, the first ten locations to have confirmed importations are those listed in Table~\ref{tab:spread_events}. The remainder of January saw cases being confirmed in several other countries. Of note is that China imposed a \emph{cordon sanitaire} in Wuhan on 23 January 2020 and that the first successful importation (in the terminology of \cite{ArinoBajeuxPortetWatmough2020}, i.e., a local transmission event) was reported by Vietnam on 24 January 2020 \cite{WHO2020_sitrep4}.
\begin{table}[htbp]
    \centering
    \begin{tabular}{l|l|l|l}
        Date & Location & Note & Source \\
         \hline
        13 Jan. & Thailand & Arrived 8 Jan. & \cite{WHO2020_first_case_outside_CHN} \\
        16 Jan. & Japan & Arrived 6 Jan. & \cite{Sim2020,WHO2020_second_case_outside_CHN} \\
        20 Jan. & Republic of Korea & Airport detected on 19 Jan. & \cite{WHO2020_KOR_2020_01_21} \\
        20 Jan. & USA & Arrived Jan. 15 & \cite{HolshueDeBoltLindquistLofyEtAl2020} \\
        23 Jan. & Nepal & Arrived 13 Jan. & \cite{BastolaSahRodriguezMoralesLalEtAl2020} \\
        23 Jan. & Singapore & Arrived 20 Jan. & \cite{CNA_2020_01_23_first_case_Singapore} \\
        24 Jan. & France & Arrived 22 Jan. & \cite{20minutes_2020_01_25} \\
        24 Jan. & Vietnam & Arrived 13 Jan. & \cite{TheHill_2020_01_23_first_case_Vietnam,NguyenVu2020} \\
        25 Jan. & Australia & Arrived 19 Jan. & \cite{first_case_Australia_2020_01_25} \\
        25 Jan. & Malaysia & Arrived 24 Jan. & \cite{NewStraitsTimes_Malaysia_2020_01_24,Elengoe2020}
    \end{tabular}
    \caption{First ten international locations having reported imported COVID-19. \emph{Date} refers to the date the case was reported. All dates are in 2020. All cases in this table were imported from China, except for Vietnam, which concerned both an imported case and a local contact.}
    \label{tab:spread_events}
\end{table}

\begin{figure}[hbtp]
\sidecaption
\centering
\includegraphics[width=\textwidth]{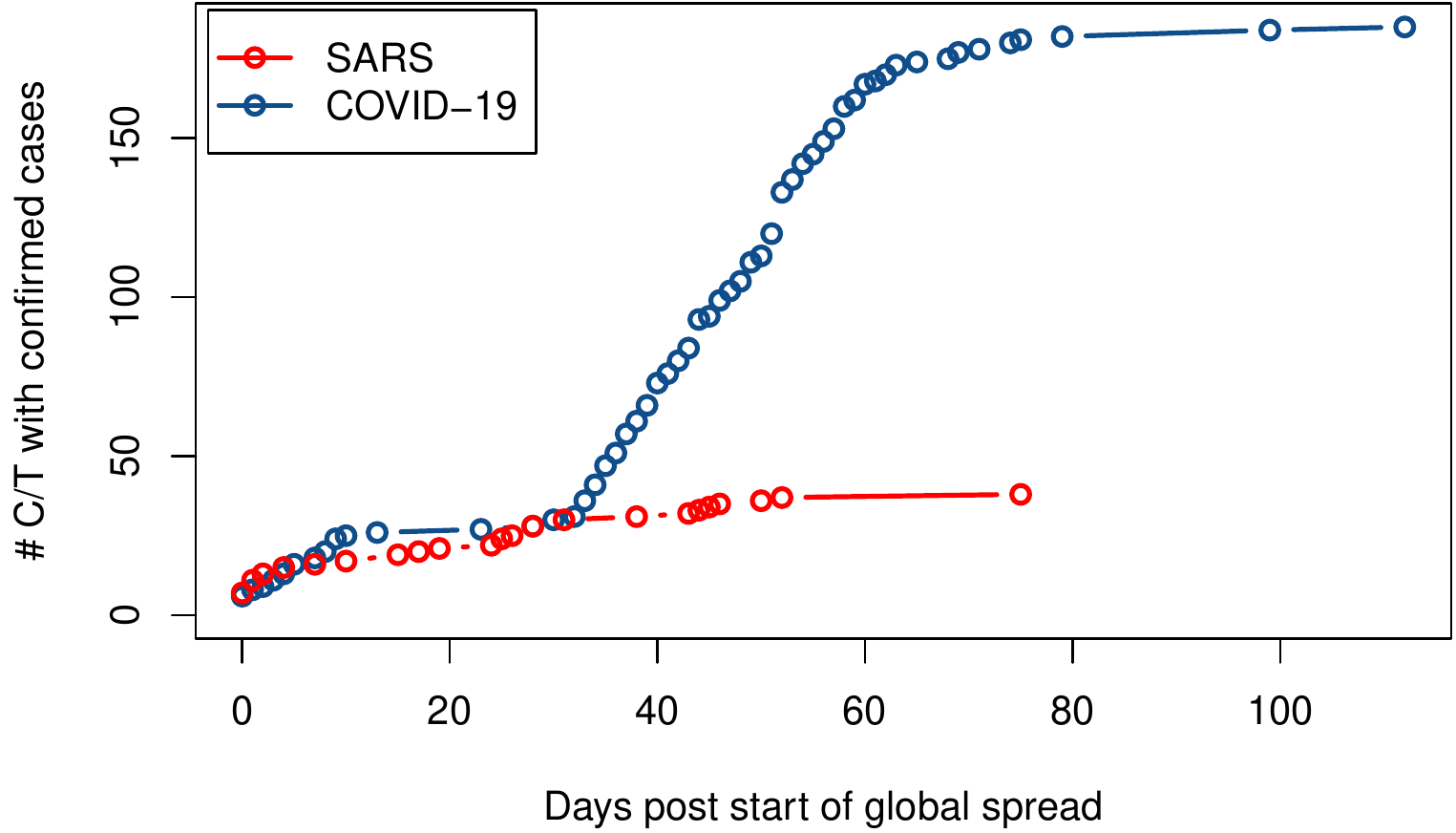}
\caption{Number of countries and territories (ISO-3166-1-alpha3 codes) having reported confirmed cases of SARS-CoV (red) and SARS-CoV-2 (blue) as a function of the number of days since importation in HKG (SARS-CoV) and importation in THL (SARS-CoV-2).}
\label{fig:nb_ISO3166}
\end{figure}

Starting in February 2020 and with more and more of the locations having reported importations earlier seeing local transmission chains, global spread accelerated. Figure~\ref{fig:nb_ISO3166} shows the number of ISO 3166-1-alpha3 codes (top level jurisdictions) reporting their first confirmed case as a function of the time since the first confirmed international exportation event.

While I do not detail them in the modelling section because of my focus there on models able to provide explanations of the phenomena, it is worth noting that interesting time series analyses were performed during the early stages of spread. 
For instance, \cite{ChristidisChristodoulou2020} used ARIMA analysis of travel data together with disease propagation data to forecast future destinations. 
The authors find that uncertainty as to the percentage of asymptomatic cases makes previsions complicated; this conclusion is in line with personal work \cite{ArinoPortet2020}.
Likewise, \cite{Hafner2020,HernandezMatamorosFujitaHayashiPerezMeana2020} considered spatial autoregressive models.
Also, although not global, continental-level spread as documented for Africa in \cite{GayawanAweOseniUzochukwuEtAl2020} is included here because the focus is not on the transition between the global level to the continental level but on spread within the continent.
Finally, \cite{MelinMonicaSanchezCastillo2020} use self-organising maps to look for similarities in epidemic curves to identify countries seeing propagation of the same type.


\subsection{Attempts to slow down the global spread}
\label{subsec:slowing_down_spread}
Using the terminology of the conceptual model of spatialisation, when COVID-19 started its international spread, there were very few jurisdictions that were exporters of COVID-19 and an immense majority of potential importers. Public health authorities in those jurisdictions that did not have cases at that point therefore took measures to try to stop or at least delay importations. To this end, they used three main types of measures: restriction or suspension of travel, entry screening and post-arrival self-isolation measures.

\begin{figure}[htbp]
    \centering
    \includegraphics[width=0.9\textwidth]{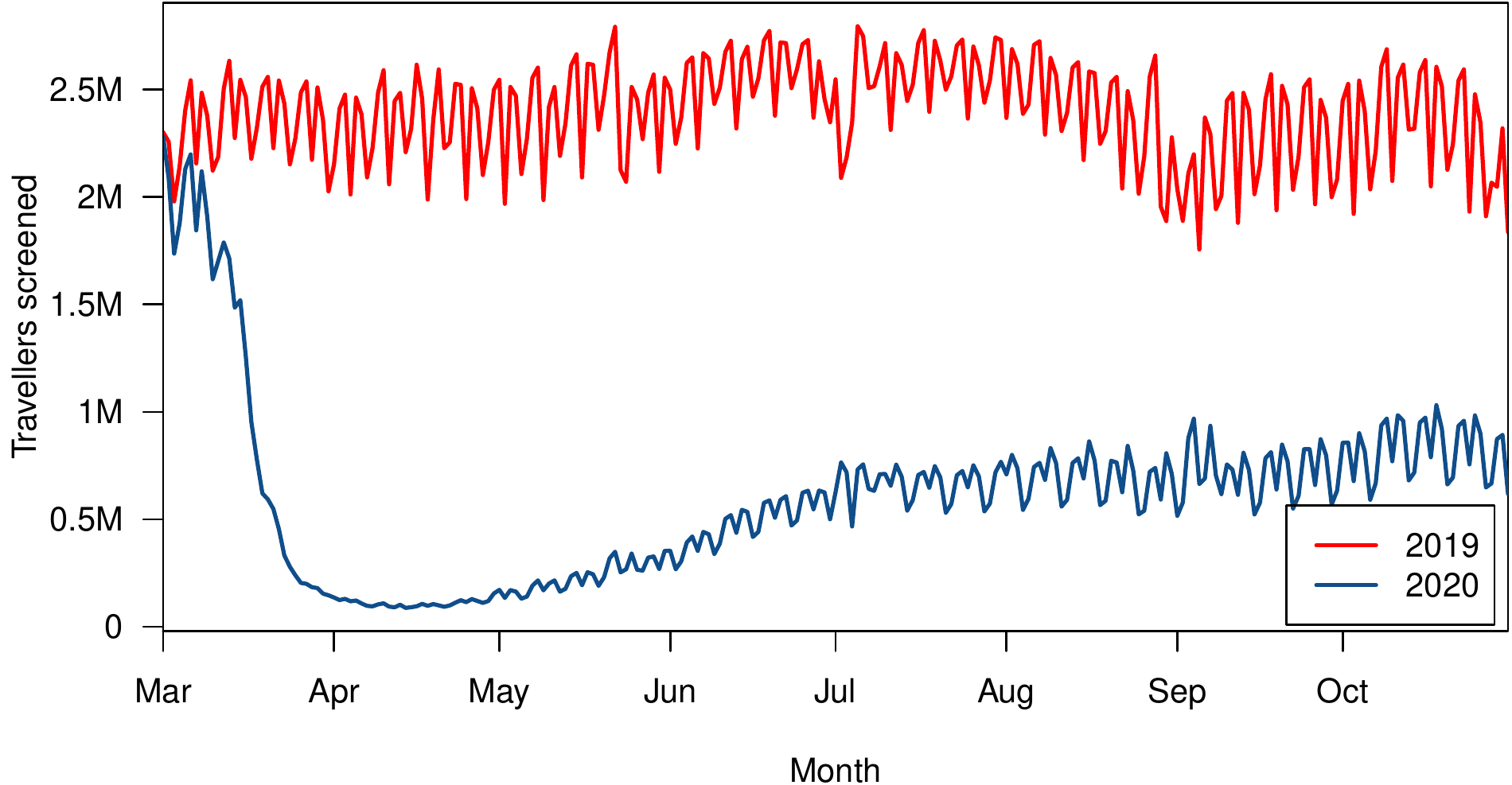}
    \caption{Daily number of passengers processed by the United States Transport Security Agency (TSA) in 2019 and 2020; data from \url{https://www.tsa.gov/coronavirus/passenger-throughput}.}
    \label{fig:TSA_screens}
\end{figure}

Starting early on and ongoing at the time of writing, various jurisdictions took measures to curtail or even interrupt travel. 
Passengers themselves also abstained from travelling.
The result of this was a precipitous drop in travel volumes.
The intensity of this effect can be seen in Figure~\ref{fig:TSA_screens}, which shows the daily number of passengers processed by the United States Transport Security Agency, i.e., the number of individuals undertaking a trip originating in the USA, in 2019 (red) and 2020 (blue).
The data shown for 2019 is for a year earlier, but shifted so it corresponds to the same day in the week.
At the lowest point, on Thursday 16 April 2020, TSA screened 3.63\% of the number of travellers they had screened on Thursday 18 April 2019.
The same trend can be observed for instance in tourism, with the United Nations World Tourism Organisation Tourism Dashboard (\url{https://www.unwto.org/international-tourism-and-covid-19}) reporting that the number of international tourists arrivals in April and May 2020 was 97\% less than the same months in 2019.

In \cite{AdigaVenkatramananSchlittPeddireddyEtAl2020}, an analysis of the global air transportation network is undertaken using the network distance defined in \cite{BrockmannHelbing2013}, attempting to tease out the effect of travel interruptions on the spread.
While both of these studies provide very interesting insights into the issue, a precise quantification of the effect of such fundamental changes to travel is hard.

Entry screening is typically implemented at ports of entry (ports, airports, border crossings) and seeks to identify individuals who are bearing the disease of concern, in order to isolate them and thereby avoid potential transmission of the disease in the local population.
There is some debate about the usefulness of entry screening, especially during the early stages of a global spread event. See \cite{MouchtouriChristoforidouAnderHeidenMenelLemosEtAl2019} for an extensive review.
The sensitivity and specificity of the thermal detection equipment used is questionable \cite{BitarGoubarDesenclos2009}. 
In the case of COVID-19, it has also been argued that this low sensitivity would combine with the fact that fever detection would often fail because of the frequency of asymptomatic cases \cite{BwirePaulo2020}.
Entry screening at the beginning of a health crisis also means looking for a needle in a haystack, since the volume of incoming passengers from all locations vastly dominates the volume of passengers coming from the location of interest \cite{KhanEckhardtBrownsteinNaqviEtAl2013}. 
Since prevalence is low at the beginning of the event, this further compounds the lack of efficacy and results in poor characteristics for the method \cite{DellOmodarmePrati2005}.
Also, screening protocols themselves vary widely from location to location \cite{GaberGoetschDielDoerrEtAl2009}, rendering a general evaluation of the value of a protocol difficult.
Despite these reservations, in the case of COVID-19, some of the evidence of early international spread comes through entry screening, so there seems to have been some limited benefit to thermal imaging entry screening.

After the initial few days during which testing was thermal imaging-based, screening switched to using much more reliable PCR tests. 
This became possible because sequencing of the virus genome was performed remarkably quickly.
As a consequence, currently, there are four main attitudes towards screening: no screening at all; ``soft'' screening, i.e., verbal or written questionnaires; testing on entry; testing prior to entry. 
Some countries use a combination of approaches, for instance requiring testing only for individuals arriving from regions considered particularly at risk.

A jurisdiction still has one option to combat the risk that successful importations take place: it can recommend or impose that individuals arriving from another jurisdiction spend some time in \emph{quarantine}.
Canada, for instance, has insisted on a two-weeks quarantine period for all incoming travellers since the beginning of the crisis, with exemptions.

\subsection{From global to local spread}
\label{subsec:global_to_local_data}

The first step in switching from a purely global vision of spread to a local one is to consider when COVID-19 could arrive in ``one's backyard''. In the early stages of the pandemic, before most top-level jurisdictions reported reported human-to-human transmission chains, it was of interest to those jurisdictions having no or few local cases to understand the risks that their connectedness to other jurisdictions carried. 
Because of evidence gathered during past pandemics and other notable public health events (see Section~\ref{subsec:drivers_of_spatial_spread}), this evaluation was mostly carried out by investigating a given jurisdiction's connection to the rest of the world by means of the air transportation network. This was the method used for instance in Mexico \cite{CruzPachecoBustamanteCastanedaCaputoJimenezCoronaEtAl2020}, India \cite{GunthePatra2020} or Europe \cite{PullanoPinottiValdanoBoelleEtAl2020}.
Much practical work on this aspect has come to rely on global airline transportation data such as that provided by the International Air Transport Association (IATA).
However, it should be noted that this dataset quickly became unreliable because of the dramatic fall in travel volumes discussed in Section~\ref{subsec:slowing_down_spread}.

\subsection{Chronology and characteristics of local spread}
\label{subsec:local_data}
Most of the very early work on local spread concerned China, since it was the first country to experience this.
The same type of method was used in \cite{FanLiuGuoYangEtAl2020} as is detailed at the start of this section, but with past data on mobility during the Spring Festival of millions of migrant workers residing in Wuhan. The aim was to assess the risk to locations visited by the migrant workers. Since Chinese New Year was on 25 January 2020, while the cordon sanitaire was imposed in Wuhan on 23 January, a lot of individuals did make the trip. This allowed the authors to venture which places were probably under-reporting cases.
See also \cite{XieQinLiShenEtAl2020}, which uses GIS techniques to study the spread within China and the factors contributing to this spread. As do the authors of the previous paper, they find that connection to Wuhan, both in terms of population flow and economically, was the main driver of the initial spread.


Tracing transmission chains originating from importations allowed to better understand the consequences of importations. 
See, for instance, \cite{BoehmerBuchholzCormanHochEtAl2020}, which breaks down such a transmission chain that started on 27 January 2020 in Bavaria (Germany).
In \cite{CandidoClaroJesusSouzaEtAl2020}, the early spread in Brazil is documented, from importation from Europe (as evidenced by genome typing of the strains) to local spread within states, finally followed by exportation from urban centres.
This is confirmed by \cite{FortalezaGuimaraesAlmeidaPronunciateEtAl2020}, who consider spread among 604 cities in S\~ao Paulo State, Brazil. They show that in the heterogeneous setting they consider, there are two patterns of spread: one spatial, where the disease spreads to the nearest spatial component; the other hierarchical, where within one unit, spread starts with the top level urban centre then makes its way to smaller cities.
In \cite{FauverPetroneHodcroftShiodaEtAl2020}, the authors use genomic and transportation data to consider the spread within the USA and conclude that quite early on in the spread, importations into uninfected locations in the country were much more likely to originate elsewhere in the country than abroad.
Propagation within the USA was also studied by \cite{HohlDelmelleDesjardinsLan2020}, in which the occurrence of space-time clusters is studied. This interestingly shows that as the epidemic took hold, there occurred more and more smaller clusters, confirming in some sense the similar observations in Brazil.
Another investigation of continental spread in the USA is carried out in \cite{MollaloRiveraVahedi2020} using a multilayer perceptron neural network. The authors use the Moran index computed on the incidence rates and a large number (57) of explanatory variables: socioeconomic, behavioural, environmental, topographic, demographic, age-adjusted mortality rates from several diseases, both infectious and chronic. They find that some of the most important factors predicting COVID-19 incidence rates are the age-adjusted mortality rates of ischemic heart disease, pancreatic cancer and leukemia, median household income and total precipitation.
In \cite{AbryPustelnikRouxJensenEtAl2020}, the time evolution in several countries and the time evolution within France are investigated using time series methods incorporating spatial components. While mostly methodological, this provides interesting tools to consider the spatio-temporal evolution of the disease across multiple jurisdictions.

Other authors considered mechanisms for slowing down the spatial spread within a country.
The authors of \cite{MishraMohapatraKumarSinghEtAl2020} advocate for a disconnection between locked-down urban centres and rural areas in India as a means to avoid complete country-wide lockdown.
This position is justified; indeed, authors in \cite{CopielloGrillenzoni2020}, for instance, found strong correlation between population density and the spread of SARS-CoV-2 in China, so it could be that forbidding movement between locations at high risk (the cities) and those at lower risk is a valid approach. However, to the best of our knowledge, no nation implemented such a system; indeed, during the initial wave, most countries implemented country-level lockdowns that also relied on severely limiting or completely interrupting mobility within their territory.
In \cite{DicksonEspaGiulianiSantiEtAl2020}, the authors consider the effect of containment measures on the spread of COVID-19 between provinces in Italy.
In \cite{KraemerYangGutierrezWuEtAl2020}, the authors used human mobility data in China to consider the spread of COVID-19 within China, in particular in relation to the impact of control measures.

\subsection{Chronology and characteristics of hyperlocal spread}
\label{sec:hyperlocal}
Hyperlocal spread was documented early on during the course of the pandemic because of cases that happened onboard cruise ships that were under quarantine. These events, while unfortunate for those involved, have provided a wealth of data. In particular, they were extremely helpful in finding out key epidemiologic parameters such as reproduction number \cite{ZhangDiaoYuPeiEtAl2020}, prevalence of asymptomatic infections \cite{EmeryRussellLiuHellewellEtAl2020}, incidence \cite{Nishiura2020}, transmissibility of the disease \cite{MizumotoChowell2020} or case fatality ratio \cite{RussellHellewellJarvisZandvoortEtAl2020}. Because cruise ships have records of who was infected together with the room they were in, it should become possible to build a good understanding of spatial aspects, although to the best of our knowledge, this data has not yet been released.

Many countries faced and are facing outbreaks in long-term care facilities (LTC). 
There are a variety of reasons for this elevated risk; see, e.g., \cite{StallJonesBrownRochonEtAl2020,TelfordOnwubikoHollandTurnerEtAl2020}. This led to trememdous effort to control such outbreaks \cite{TelfordOnwubikoHollandTurnerEtAl2020}.
Movement within LTC can be documented (and modelled) accurately; see, e.g., \cite{ChampredonNajafiLaskowskiChitEtAl2018,NajafiLaskowskiBoerWilliamsEtAl2017}. 
The health of residents is also monitored (usually) well. 
As a consequence, nosocomial COVID-19 outbreaks also provide valuable data at the hyperlocal level.
See documented outbreaks in \cite{IritaniOkunoHamaKaneEtAl2020,KuhnRose2020,McMichaelClarkPogosjansKayEtAl2020,ShiBakaevChenTravisonEtAl2020,ShraderAssadzandiPilkertonAshcraft2020}.

Note an interesting ``twist'' on hyperlocal spread: in \cite{HoltzZhaoBenzellCaoEtAl2020}, the authors conduct a wide-ranging analysis at the hyperlocal scale, in the sense that the consider the movements of individuals at the local scale but over the entire territory of the United States of America. This allows them to consider the effect of spatial heterogeneity of public health orders.

\section{COVID-19-specific models}
\label{sec:covid19_modelling}
I replicate here the hierarchical spatial structure in Section~\ref{sec:chronology_and_characteristics} rather than the methodological one in Section~\ref{subsec:spatial_models}. 
Indeed, while most work detailed here falls within one of the three classes of methods in Section~\ref{subsec:spatial_models}, I also report on other methods that gave interesting results.

\subsection{Models of global spread}
\label{subsec:global_spread_models}
To the best of my knowledge, most models for the global spread had in their objectives to study how to slow down the global spread of the infection or considered spread within specific countries or groups of countries; these are discussed in the relevant later sections.

In \cite{ArinoPortetBajeuxCiupeanu2020,ArinoPortetRees2020}, we set the $SL_1L_2I_1I_2A_1A_2R$ model of \cite{ArinoPortet2020} in a metapopulation context and focused on the risk of importation in different countries. The model was run daily to provide the Public Health Agency of Canada with an assessment of the most likely countries to import the disease in the coming days. The model includes travel at different levels, which, as pointed out in Section~\ref{subsec:global_to_local_data}, was a documented feature of spread.

In \cite{SiwiakSzczesnySiwiak2020}, an SIR-type metapopulation model in the GLEAM framework \cite{BalcanGoncalvesHuRamascoEtAl2010} is used that combines population densities, commute patterns and long-range travel. Used at the early stage of the spread, the authors find that it is likely that the value of the basic reproduction number $\R_0$ and the prevalence are badly estimated in some locations, with estimates in the literature at the time driven by locations with a large population. They conclude that the number of cases was probably underestimated.

\subsection{Modelling the slowing down of global spread}
\label{subsec:slowing_down_spread_models}
In \cite{ChinazziDavisAjelliGioanniniEtAl2020}, a metapopulation model for the global spread of COVID-19 is used to consider in particular the role of international travel bans. The authors show that while the cordon sanitaire in Wuhan did little to slow spread within China, its impact internationally was more pronounced. The combined effect of travel restrictions and community effort is also studied, with the interesting finding that travel restrictions alone do not suffice to have an effect on propagation.
In \cite{AdekunleMeehanRojasAlvarezTrauerEtAl2020}, a stochastic SEIR metapopulation model is used, together with Official Airlines Guide (OAG) data, to consider the role of travel restrictions taking place after 24 January 2020. The authors found good adequation with the number of imported cases in several countries as of the end of January. They focused in particular on Australia and establish that the travel ban there might have delayed the onset of widespread propagation by four weeks.

A stochastic simulation model is used in \cite{DickensKooLimSunEtAl2020} to consider different scenarios regarding testing (rather than screening) of incoming individuals and the duration of quarantine periods. 
Similarly, \cite{CliffordPearsonKlepacVanZandvoortEtAl2020} use a stochastic model to quantify the effectiveness of screening and so-called \emph{sensitisation} of travellers, i.e., the provision of health information in an effort to trigger compliance with self-isolation recommendations.
In \cite{RibeiroCastroESilvaDattiloReisEtAl2020}, the example of air transportation in Brazil is considered using an SIR-type metapopulation. The speed of spread in relation to network measures such as centrality was explored, with closeness centrality shown to be a good predictor of the vulnerability of a city.

In \cite{ArinoBajeuxPortetWatmough2020}, we considered the risk of disease importation in a location that is seeing little to no local transmission chains. 
As with most of our work on the subject, we used a modified version of the model in \cite{ArinoPortet2020}. 
In this case, we used a stochastic version, which we subjected to stimulations to represent the inflow of infected individuals into a location.
The model also allowed us to quantify precisely the effect of quarantine in terms of its effect on the inflow rate.

\subsection{Modelling the transition from global to local spread}
\label{subsec:global_to_local_spread_models}
In \cite{GilbertPullanoPinottiValdanoEtAl2020}, the risk of importation of COVID-19 in African countries was considered using air travel data as well as data from the Monitoring Evaluation Framework (MEF) of the WHO International Health Regulations. The model is quite simple and comprises no dynamic components, meaning that it provides a snapshot evaluation of the risk of importation. As it was formulated at the beginning of the spread event, when most of the exportation was assumed to come from China, it does nonetheless provide meaningful results.

In the already cited \cite{ArinoBajeuxPortetWatmough2020}, we focused on the risk of importation of COVID-19 in locations that are seeing little to no local transmission, thereby considering the interface between the rest of the world and such locations. We showed that the probability of importation was most dependent on the rate at which cases are imported in the locations, but that the outcome of a successful importation was then determined to a large extent by the intensity of public health measures in the locations.

\subsection{Models of local spread}
\label{subsec:local_spread_models}
The location for which data became readily available the soonest was China. Since China is also a very large country, some very interesting work was carried out in the context of spread within that country.
The authors of \cite{WuLeungLeung2020} considered the spread of COVID-19 within China using an interesting idea: they estimated the size of the outbreak in Wuhan from known international exportations, then used a metapopulation model with Wuhan as the source of infection to estimate spread within China. 
In \cite{LiPeiChenSongEtAl2020}, the role of undocumented infections is investigated in relation with the spread of COVID-19 between 372 Chinese cities using a metapopulation SEIR model incorporating documented and undocumented infections.
In \cite{HuangLiuDing2020}, spread within Hubei Province and in the rest of China is investigated using statistical tools (the Moran index and a logistic model). Then an ODE SEIR model is used to compute $\R_0$ in the different locations.
In \cite{SunXiaoJi2020}, an SEIR-type model with additional compartments for diagnosed and confirmed, suspected and infected as well as suspected but uninfected individuals is set in a metapopulation framework with two patches: Hubei Province and the rest of China. The model is used to consider the effect of lifting lockdown measures.

In \cite{Amar2020}, a simulation platform is used to consider the spread in France. The model operates at the level of subregions (\emph{d\'epartements}) and supposes that individuals can be susceptible, asymptomatic, symptomatic, recovered, hospitalised and diseased. An interesting feature of the paper is a comparison between the results of continuous time deterministic and discrete stochastic methods, with the latter showing better adequation with observed data.

The authors of \cite{CandidoClaroJesusSouzaEtAl2020} considered spread of SARS-CoV-2 within Brazil. This colossal endeavour considers actual genotyping of the virus and prior to modelling work proper, details importations and the spatio-temporal spread of various genomes of the virus. Spatio-temporal modelling then uses a continuous phylogeographic model. The model is not predictive but sheds light on the spread process: they find that spread was mostly local, i.e., within state borders. Both within-state and between-state spread was also found to have decreased after the implementation of NPI.

In \cite{AlthouseWallaceCaseScarpinoEtAl2020}, the effect of heterogeneity of policies in the USA is investigated. A model is formulated that is a metapopulation in essence; based on data on people movement to places of gathering such as churches, the model allows the redistribution of individuals between locations following different types of policies. They observe that spatial heterogeneity in measures tends to increase the likelihood of subsequent infection waves. Spatial heterogeneity is also investigated in \cite{CuadrosXiaoMukandavireCorreaAgudeloEtAl2020}, which uses a metapopulation model to probe the impact of disparity of healthcare capacity in Ohio.
In \cite{ChangKahnLiLeeEtAl2020}, the effect of changing travel rates within and between locations is investigated, with data for Taiwan.

Finally, note that because COVID-19 is spreading globally and that national level jurisdictions (and sometimes even lower level ones) implemented a variety of responses, it is useful to compare the situation in different jurisdictions.
Even though this is not spatial modelling \emph{stricto sensu}, such works are worth mentioning here as they provide the underpinning to spatial models. 
In \cite{HeChenJiangJinEtAl2020}, the authors use an SEIR model to compare transmission patterns in China, South Korea, Italy and Iran.
In \cite{HauserCounotteMargossianKonstantinoudisEtAl2020}, an age-structured SEIR model is used to compare the dynamics of disease spread in Hubei Province and six European regions. The focus is on the estimation of the case-fatality (CFR), symptomatic case-fatality (sCFR) and infection-fatality (IFR) ratios. The authors find that the latter two indicators are better suited to describe the potential impact of the pandemic and note that they find geographic heterogeneity of the estimated values. This heterogeneity is not only between Hubei Province and the European locations under consideration, but also between the European locations themselves.
With collaborators, we used the model in \cite{ArinoPortet2020} to provide a daily forecast of spread in several Canadian provinces \cite{ArinoPortetBajeuxCiupeanu2020} and found that estimates for some parameters were consistent across provinces while estimates for others varied widely, in particular, the proportion of asymptomatic cases.

\subsection{Models for spread at the hyperlocal level}
\label{sec:hyperlocal_models}
A lot of work during this pandemic has focused implicitly on the hyperlocal level, but recall that here the object is models in which I found an explicit reference to spatial aspects.

In \cite{JennessWillebrandMalikLopmanEtAl2020}, a network model is used to model the spread of SARS-CoV-2 onboard the Diamond Princess cruise ship, with nodes representing individual passengers and crew members. 
Age-structure was used as well.
The model was calibrated to known transmission data and the effect of control measures was then considered.
See also the already cited \cite{EmeryRussellLiuHellewellEtAl2020,MizumotoChowell2020,Nishiura2020,ZhangDiaoYuPeiEtAl2020} for more modelling work related to spread aboard the Diamond Princess.

The authors of \cite{BraunTaraktasBeckageMolofsky2020} used an SLIAR agent-based model to consider the effect of social distancing, viral shedding and what they call the social distance threshold. They find that the three lead to threshold behaviour (``phase transitions'') that have different effects on the course of the epidemic.

In \cite{FioreDeFeliceGlicksbergPerlEtAl2020}, ABM are used to consider in particular the effect of testing policies. Agents are distributed on a map depending on the population density in the areas under consideration. They are also assigned movement patterns that can cover the whole map, a medium range or a small one. Some interesting observations are that when tests have low reliability or that the ability to trace contact is low, a large fraction of the testing capacity remains unused despite an increasing incidence. They also find that mixed testing policies are useful to contain spread.


\section{New variants}\label{sec:new_variants}
SARS-CoV-2 is an RNA virus and as such is subject to high mutation rates leading potentially to variants \cite{DuarteNovellaWeaverDomingoEtAl1994,KautzForrester2018}. Thus the emergence of new variants was expected from the onset of the crisis.
At this point, there are several major variants to the original variant that have been detected. 
This number can be expected to rise: detection of most variants requires genome sequencing, which is performed at different rates in different countries \cite{Furuse2021}, meaning that capacity to detect variants varies greatly globally.
Of particular interest at the time of writing is B.1.1.7, which was first detected in the United Kingdom in early December 2020 but is presumed to have been spreading since as early as September 2020. This variant is particularly concerning as it appears to be more transmissible than the original variant. 
It seems that this variant should not, however, be detrimental to the ongoing vaccination efforts \cite{ContiCaraffaGallengaKritasEtAl2020}. 
Many countries took preventive measures in order to delay the arrival of the variant, essentially forbidding all travel originating from the United Kingdom, but given that circulation probably started several months before these measures, their efficacy is debatable. 
For instance, \cite{DuWangYangAliEtAl2021} estimates that, of 19 countries evaluated, 16 had at least a 50\% chance of having already imported the variant by 7 December 2020.
The novel variants led some countries to consider exit control measures; some European Union (EU) countries (Belgium and France, for instance) decided late January 2021 or early February 2021 to forbid both entry from and exit to non-EU countries for non-essential travel. 
In the context of pandemic H1N1 influenza, exit screening was shown to have the potential to be more an efficacious control measure than entry screening \cite{KhanEckhardtBrownsteinNaqviEtAl2013}. 
It is therefore interesting to see this type of control finally being applied, although the intent is not the one we were advocating in \cite{KhanEckhardtBrownsteinNaqviEtAl2013}.

Modelling the spatio-temporal spread of these novel variants can be conducted in very much the same manner as was done for the original variant. For instance, metapopulation models for multiple species such as those considered in \cite{ArinoDavisHartleyJordanEtAl2005,ArinoJordanDriessche2007} can be readily adapted to a multiple variant situation. However, it is important to bear in mind that because of the detection issues mentioned earlier, these models are hard to parametrise when considering the initial spread of the variants.

\section{Discussion}
This is but a brief and very incomplete snapshot of the state of knowledge about the spread of COVID-19 at the time of writing in December 2020, with a few additional details about the new variants added in January 2021.
As indicated, it is likely that I omitted a lot of publications on the subject, given the immense amount of literature COVID-19 has generated.

From the perspective of the spatio-temporal spread of the disease, although there is still much to learn, I think we also now have the \emph{luxury of hindsight}: many groups, mine included, have produced a variety of models in the first few months of the crisis, which can and should now be confronted to the reality of the outbreak.
Because COVID-19 is so widespread, there is less urgency to consider its spatial spread in the perspective of emergency response and the focus could now evolve, at least in part, to the evaluation of the models we produced. The problem of re-importation of the disease in locations having managed to drive it away remains an important one, so I am not advocating to stop all work regarding spatial spread; I am only pointing out that understanding what worked and what did not during the initial spread would actually help for these subsequent importation events.

Going forward, though, I believe that there is still a lot to be done on one key aspect of spatio-temporal models: most of the work carried out by those of us working in this area has come to rely on one particular dataset, the so-called IATA air transport data. Figure~\ref{fig:TSA_screens} shows that in the particular instance of COVID-19, the quality of this data leaves a lot to be desired. When the data for 2020 becomes available in 2021, it will be extremely important to scrutinise it in order to understand what type of changes took place. 
Another important point to ponder will be the use of other data sources to compensate for this loss of relevance of IATA-type data.
Cell phone location data is showing promise, but it suffers from several limitations, the most important of which being that it is most useful and detailed at the country level and, more importantly, that it is either proprietary or extremely expensive to acquire.

\begin{acknowledgement}
I am supported in part by NSERC and by CIHR through the Canadian COVID-19 Mathematical Modelling Task Force.
I acknowledge support both financial and logistical from the Public Health Agency of Canada.
\end{acknowledgement}


\end{document}